# A Simple Reactive-Flow Model for Corner-Turning in Insensitive High Explosives, Including Failure and Dead Zones. I. The Model.

Peter Todd Williams*[a]

**Abstract:** I report on a novel simple reactive-flow model that captures corner-turning behavior, including failure and the creation of dead zones, in insensitive solid heterogeneous high explosives. The model is fast, has a minimum of free parameters, and is physically grounded in the underlying initiation and burn phenomena. The focus of the model is explosives based on triaminotrinitrobenzene (TATB), although the model is quite general. I initially developed the model and integrated it into a branch of a Lawrence Livermore National Laboratory (LLNL) arbitrary Lagrangian-Eulerian (ALE) code concurrently with other modelling efforts of mine, but I subsequently adopted it to study corner-turning behavior in the TATB-based high explosive PBX 9502 when two other available models suggested to me proved inadequate, being either prohibitively computationally expensive or numerically unstable. An early version of the model reproduces corner-turning behavior in two different double-cylinder tests and a mushroom-type corner turning test. The model was informed and influenced by other reactive-flow models applicable to shock initiation or corner-turning, most especially including JWL++ Tarantula 2011 and an in-house piecewise-linear CHEETAH-based model that exhibits corner-turning behavior, but also, to a degree, by Ignition & Growth (I&G), CREST, and the Statistical Hot-Spot Model. The model also shares some similarities to SURF models. In this paper, I describe the model and its theoretical development as I originally conceived it, as well as subsequent improvements.

## 1 Introduction

For design purposes – that is to say, for the purposes of designing precision munitions and other devices – the simplest model for the steady or nearly-steady detonation of high explosives (HE) rests on Chapman-Jouguet (C—J) theory [1, 2], which requires not even a full equation of state (EOS) for the detonation products, but merely the adiabat for the isentropic expansion of detonation products behind the detonation wave. In principle, this theory is only applicable to detonation shock waves that are not curved, but in practice, for most conventional HE (CHE), the reaction zone (RZ) is thin enough relative to the radius of curvature of the detonation front in almost all regions, and for almost all applications, such that this restriction proves not to be too cumbersome. One simply assumes that detonation being initiated at some point, a detonation wave spreads outwards at a uniform speed from the front following Huygen's principle; this is sometimes known as *programmed burn* [3–5]. The situation changes dramatically however for insensitive HE (IHE), in which the thickness of the reaction zone may be non-negligible relative to the curvature of the detonation front or indeed even relative to design features in various devices, and in which the shock initiation and transient rise to detonation are also non-trivial. Practical theoretical and numerical treatment of the detonation of IHE demands the development of more sophisticated models and theories than the C—J theory on which programmed burn rests. The quintessential IHE is triaminotrinitrobenzene (TATB); TATB is the explosive component of the formulations PBX 9502 and LX-17 and was the focus of the work that led to the development of the model presented in this paper.

The simulation of IHE in complex geometries challenges theory. It demands a much broader understanding of the detonation process than is required for quasi-1D detonation. Currently, there are two predominant theoretical approaches to the numerical modeling and simulation of the detonation of IHE in complex geometries. One theory, the Detonation Shock Dynamics (DSD) theory [6–8], is based on the notion that the detonation shock speed can be written as a function of the curvature of the shock front. Known as the *curvature effect*, this is a very good approximation in most instances. However, while DSD has a certain elegance, and certainly has computational economy, it has difficulties in problems that exhibit failure, or when the RZ thickness is comparable to the radius of curvature of the shock front. Among the US nuclear weapons laboratories at least, DSD is the dominant theoretical approach taken at Los Alamos National Laboratory (LANL). The other predominant approach, and the focus of this paper, reactive flow modeling (RFM) [9–14], attempts to solve the reactive fluid equations for detonation. This is done with the advantage of grossly simplifying assumptions of course, but in a manner such that the detonation wave emerges naturally rather than having to be put in explicitly, "by hand," so to speak. RFM is thus analogous to so-called shock-capturing methods in computational fluid dynamics (CFD), in contrast with shock-fitting methods, to which DSD is correspondingly analogous. The RFM approach is much more demanding of computer resources than DSD, but in principle, it promises to be more general; it is the predominant approach taken at Lawrence Livermore National Laboratory (LLNL).

[a] *P. T. Williams*
*NorthWest Research Associates*
*301 Webster St., Monterey, CA, USA*
*\*e-mail: peter.todd.williams@protonmail.ch*
*ptw@nwra.com*

Conceptually, the RFM approach divides the problem of simulation of the detonation of IHE into two components. First is the problem of the equation of state (EOS), that is, the equation of state of both the reacted and the unreacted explosive, as well as the problem of the EOS of the partially-reacted explosive; I will return to this later. The second part is a rate model. In principle, one could imagine any number of complex coupled rate equations. In reality, of course, the actual chemical reaction network in the detonation of a modern solid plastic explosive is horrifically complicated, even if it were possible to learn it in the first place. In practice, what is often done – and what is done here – is to assume that the reaction progress can be described by a single parameter $F$; the fraction $F$ of the HE is assumed to be fully reacted or "burned" and obeys one EOS, and the fraction $(1-F)$ is assumed to be completely unreacted and follows a separate EOS. The rationale for this is that, in the reaction zone behind the detonation front itself, on a small scale of order the grain size or smaller, the reaction proceeds like deflagration on small subsonic burn-fronts, and each of these small burn fronts can be assumed to be infinitesimally thin. One can imagine, for example, unburned solid grains separated from burned reaction products by a negligibly-thin surface layer. The actual chemical reactions themselves are many orders of magnitude faster than the reaction-zone timescale. Even if one could resolve the individual reactions, there would be no point to it. The reaction rate is controlled by the spatially-heterogenous nature of the burn, not by bulk chemical kinetics. This comports with the present understanding that the detonation of practical heterogeneous (as opposed, say, to monocrystalline) solid HE takes place by initiation of detonation at many small defects or other "hot-spot" initiation sites such as on grain surfaces or interior to grains at defect locations.

In principle, the RFM approach should be ideally-situated to study what is known as the corner-turning (CT) problem [15]. In CT, the detonation shock wave must make – or rather, one should say, is given an opportunity to make – an abrupt turn, such as a sharp ninety-degree turn in the experiments of concern here. While this can be studied with DSD, it is a challenge, as CT pushes DSD to its limit. The CT phenomenon includes small radii of shock curvature as well as potential for failure, both of which are problematic for DSD, whereas RFM does not suffer from these limitations, at least not to the same degree.

Although such a problem is difficult to simulate well, there is ample motivation to do so, as the CT problem is an important one to solve for designers and other consumers of HE models. It is thought to be manifest in essentially all explosives, including TNT and RDX [16,17], but in practice, it is most acute in IHE and in TATB in particular, including the TATB-based explosives PBX 9502 and LX-17 [15,18–21], subjects of much current and ongoing research for their practical applications. Understanding CT in such important materials is therefore of great current and continuing interest.

While the CT problem appears in a broad range of complex and interesting geometries, the phenomenon has come to be best studied and understood in a set of classic 2D axisymmetric experiments, including double-cylinder tests and mushroom tests. In the double-cylinder test, a cylinder of HE of one diameter, $D_1$, is butted directly end-to-end (that is, flat-to-flat) against a cylinder of a second, larger diameter $D_2$, such that both cylinders share a common axis of symmetry, thus preserving axisymmetry for the overall system (see Figure 1). A detonation wave is initiated at the distal end of the smaller-diameter cylinder and proceeds along that cylinder towards the larger-diameter cylinder. Effectively, viewed in 2D, the cylindrical radius makes a simple step-function-like jump from $D_1/2$ to $D_2/2$. The detonation wave, if it is to detonate all of the HE in the second cylinder, must, again, make an abrupt turn of (nearly[1]) nine-ty degrees. One of the standard double-cylinder tests is S-COT (or, Pantex corner-turning test) [22]; a more recent version, requiring less HE, is known as ECOT [23]. In a similar class of experiments, the second (larger) cylinder is replaced with a hemisphere, making a composite object that is roughly mushroom-shaped [24]; in the case of the recently-developed test SAX [25] first constructed at LLNL, the hemisphere of HE is capped by a thin witness shell of metal for photon Doppler velocimetry (PDV) probes.

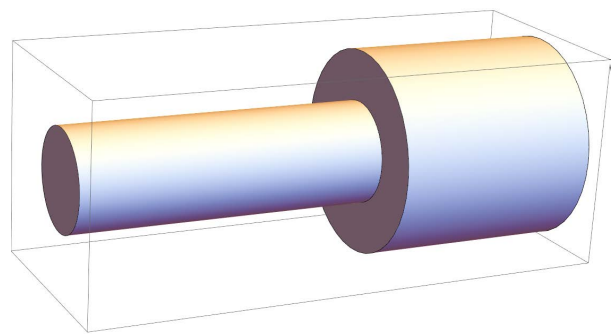

**Figure 1.** Generic conceptual geometry for double-cylinder test. Detonation wave enters donor cylinder (or rate stick) from left and proceeds to the larger-diameter acceptor cylinder (right).

---

[1] One could quibble that, due to front curvature in the donor, the required turn is slightly less than ninety degrees.

Generally, both for double-cylinder as well as mushroom-type tests, the detonation wave fails to make the instantaneous ninety-degree turn, at least for TATB in these standard tests. The detonation then leaves behind a certain amount of HE corresponding to the roughly doughnut-shaped region of HE that is left un-detonated. This region is known as a dead-zone (DZ). See, for example, Figure 3 of [22]. Experimenters in this case may have the strange task of having to scavenge powder or even chunks of solid un-detonated HE from the test-stand area after a successful firing. In more extreme cases, the detonation wave may simply bore directly through the second block of HE, effectively making an "apple-core" in the main charge. In the worst case, the detonation wave may die out completely, leaving at most only a large pit in the base of the second cylinder (or hemisphere). Outside of complete failure or even apple-coring however, the mere presence of any dead-zone may itself be of concern if the size of the dead-zone is not highly reproducible from shot to shot, because a large variation in the size of the dead-zone implies a large variation in the total chemical energy delivered by the HE. For obvious reasons, this is potentially of immense practical concern for design purposes, if that energy needs to be delivered to something else. It is therefore important in this case that factors leading to variation in the size of the dead-zone may be elucidated by theory working in concert with experiment, so that theory might have some hope of developing a model with some predictive capability, capability which could then be used by designers studying integrated experiments and devices that may be subject to similar effects.

Towards this end, there have already been several successful modeling efforts devoted to the use of RFM models to capture CT behavior and the creation of dead zones. For example, the Ignition and Growth (I&G) model has been shown, using DYNA2D, to capture CT behavior in S-COT [26]. The SURF model [12], implemented in the LANL code Pagosa, has reproduced DZ formation [27], and CREST [10,11] has captured CT behavior in PBX 9502 [28] in a double-cylinder test. Simulation of CT behavior, including DZ, has also been claimed by other researchers as well [29,30]. It has additionally been seen in the two LLNL models described below.

I sought to study lot-to-lot variations in CT and DZ behavior in batch-produced PBX 9502 in a suite of select experiments. The two models that were immediately available to me, and which were suggested to me, were a recent JWL++-based model, and a model based on a piecewise-linear rate law and EOS implemented in `CHEETAH 8.0`. (Note that `CHEETAH` is a thermochemical code that solves the EOS of the HE reaction products by minimizing the free energy subject to the thermodynamic state constraints, *e.g.* pressure and density [31]. This requires, as input, various parameters characterizing the self-interaction potential of all species of interest, as well as approximations and assumptions for the interaction energy of unlike species.) Both models ran in an (unnamed) ALE code at LLNL that is widely-used for reactive-flow simulations, dynamic simulations of HE and metal, and other applications. Regarding the model that used `CHEETAH` linked to this code, previous work at LLNL had reproduced CT behavior in S-COT with a rate model developed and implemented in `CHEETAH` that used, among a number of other ingredients, a piecewise-linear rate law called through the `CHEETAH` subroutine `pqplf2`. (Hereafter, I refer to this class of piecewise-linear models in the LLNL ALE code with linked `CHEETAH` as VPLC.) This seemed to perform well in terms of fidelity but ran quite slowly. This was my first option. In addition, although the original JWL++ model [13] had difficulty with CT, so-called Tarantula V1 [32] and, later, JWL++-Tarantula 2011 (hereafter, JT11) [33], were able to capture CT behavior in S-COT. JT11 was my second option.

In the course of my work however, it became apparent that these two available models presented certain insurmountable difficulties and obstacles, most notably regarding execution time and stability. For example, the use of a piecewise-linear rate law through a `CHEETAH` subroutine in VPLC, while it may offer certain advantages that may be appropriate elsewhere, also requires substantial computer time. This is in part because an individual run using `CHEETAH` for the EOS is much more demanding of computational resources than would be a run using a far simpler JWL EOS. The model is also all the more demanding of computer resources, however, because a model with a large number of free parameters requires a much larger number of runs in order to be adequately explored and fit to available data.

By comparison to VPLC, JT11 offered much faster execution time (approximately 20x faster than VPLC) and had fewer free parameters. I encountered, however, some unstable behavior with this model, and I suspect that hints of this instability may be present in Figures 1b, 8b, 9b, and 10d of the key reference for JT11 [33].

Fortunately, I was able to fall back upon a novel phenomenological model that I had developed earlier, the model I herein name `salinas 1.0`. The focus of this paper is the introduction of that model, which I found to satisfy my acute modeling needs to study corner-turning and the formation of dead-zones in a few select experiments. Unfortunately, however, as I no longer have access to any of my results demonstrating the success of this model, I can only describe the model and its development and subsequent improvements in this brief report, leaving demon-stration of the success of the model, I hope, to a future paper, either by myself or by others.

## 2 Model Goals

The immediate task that inspired this work was, again, the development of lot-specific models of PBX 9502, focusing on corner-turning including failure. In general, for good modelling practice, I expect a model such that:

a) The model should be able to fit a range of different experiments fairly well (in this case focusing mainly, but not exclusively, on corner-turning).
b) The model should do so with the same settings for the model parameters for all experiments.
c) The model should have as few free parameters as possible.
d) The model should be physically-motivated; it should not just be an empirical fit with ad-hoc functional forms under the hood.
e) The model, as implemented, should be minimally computationally expensive, and numerically stable.

It is worth discussing (b) in particular. I and others have found, for example, that a lot of PBX 9502 that underperforms in corner-turning, as measured by ECOT or S-COT, can be better-fit in simulation by decreasing the rate by hand, *i.e.* reducing $dF/dt$ by an overall multiplicative factor. Unfortunately, in DAX, that same lot will typically over-perform, accelerating the metal disk more than the reference lot. One can bring simulation of DAX back in agreement with experiment by adjusting the rate, but this time, one has to *increase* it. This serves as a warning that, in the end, the concordance between theory and experiment can be made to look quite impressive for anyone experiment [14], but such an exercise has limited scientific value or utility. What is needed is a model with predictive capability; a high-quality fit to a single experiment does not suffice.

Regarding the specific physical phenomena of interest for my model, note that I decided not to include any shock desensitization (also known as dead-pressing) in the model. It appeared that such desensitization was not a requisite ingredient in any of the experiments under consideration, and so, in line with my simplicity criterion, I did not include this effect.

Finally, as a practical matter, note that, as I was working with a range of development lots with different properties and could not assume a calibrated EOS from the outset, I required a separate type of experiment besides just CT experiments alone. This is because in practice, focusing only on CT experiments, some of the effects of EOS and rate-law become at least partly co-mingled. Without a dedicated EOS experiment, it can be difficult or even impossible to resolve this ambiguity.

## 3 Experiments

Initial development, testing, and fitting were performed on nominal legacy data from reference-lot tests with S-COT. A later development, testing, and fitting of the model, particularly including fitting to a range of lots of interest, required that I and my co-workers settle on a few key experiments for which we had data for the relevant production lots. Those experiments, described below, include the CT experiments ECOT and SAX, plus an experiment to fix the EOS.

Ordinarily, the EOS would be fixed by a standard cylinder test (CYLEX) [34]. Data for development lots from CYLEX available at LLNL was limited, however, due to the expense of that test and its large demands on quantity of HE for material under test (MUT), which can be problematic for small development lots. I and my co-workers, therefore, opted for data from DAX [35], which, while somewhat lower-fidelity than CYLEX, is much less costly, can be performed at higher turn-around, and most importantly is less demanding on quantity of MUT. The focus of this paper is a model, and I do not present the lot-specific fits I performed to ECOT, SAX, and DAX, nor my fits to nominally-performing lots in S-COT, but it is worth describing all of these experiments briefly for context. A more extensive discussion of the experiments can be found in the references provided.

The S-COT experiment (also known as the Pantex corner-turning experiment, as it was developed at Pantex [22]), consists of an RP-1 detonator feeding into one 25.4 mm × 19.05 mm OD pellet of PBX 9404 booster, which then impacts four 25.4 mm × 19.05 mm OD pellets of MUT, which form the "rate-stick" or donor charge. The acceptor charge consists of one large 50.8 mm × 50.8 mm cylinder of MUT. The MUT is typically either LX-17 or, in our case, PBX 9502. Data consists of streak-camera records of shock breakout on the sides of the acceptor (Figure 2).

ECOT, the Enhanced COrner-Turning test, was developed at LANL [23] as an alternative to S-COT. Following the detonator and booster, it consists of a donor, or rate stick, of five end-to-end 8 mm × 9 mm OD pellets of PBX 9501. The acceptor is two 25.4 mm × 25.4 mm OD pellets for an OAL of 50.8 mm for the acceptor. The assembly is held fixed by 3D-printed plastic judiciously-placed so as not to provide any significant confinement for the HE (Figure 3). The data of interest consist, as for S-COT, of side-breakout streak-camera traces. For my interests, MUT, as for SAX and DAX below, was specific small-batch lots of PBX 9502. ECOT is much less demanding of quantity of MUT and is less expensive than S-COT. On the other hand, the PBX 9501 rate stick is close to failure, meaning that its front-curvature is both substantial and more variable from shot-to-shot than one might desire. In the view of some researchers at LLNL, this may negatively affect shot-to-shot repeatability of overall ECOT results, as the CT behavior of the acceptor is highly sensitive to variations in its initiation by the donor.

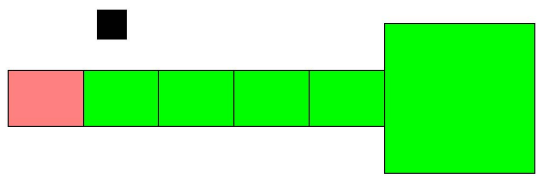

**Figure 2.** S-COT geometry (longitudinal section). Black square is 1 cm × 1 cm. Pink region is booster; light green is MUT (*e.g.* PBX 9502). Support structure not shown.

SAX is a novel modified mushroom-style CT experiment developed at LLNL [25]. Following an RP detonator and a die-pressed 12.7 mm x 12.7 mm OD booster of, *e.g.* PBX 9404 or LX-10 (HMX), the donor consists of five die-pressed 12.7 mm x 12.7 mm pellets of MUT for the "rate stick" input charge. The rate stick has light confinement by 3D-printed plastic supporting assembly, with 4 piezo pins to measure detonation speed. The acceptor charge is a 50.84 mm OD hemisphere of MUT. In a departure from older mushroom-style tests, the MUT is encapsulated by a 50.94 mm ID high-tolerance witness shell of 0.8128 mm (0.032″) thickness AL1100. The witness shell is extended by an 8.49 mm-long cylindrical skirt and is capped by a domed plastic PDV holder. For initial development work, the plastic was 3D-printed VeroWhite Plus FullCure835. Data consists of PDV traces taken at a choice range of angles away from the axis of symmetry. Outside of the "North Pole" PDV, all PDV probes appear in triplicate, spaced out by 120° around the axis of symmetry (Figure 4).

DAX [35], the experiment fixing the EOS, consists of an RP-80 detonator, followed by a 25.15 mm OD plastic holder for a 12.7 mm OD train of pellets, starting with PBX 9407 booster, hitting three 25.4 mm-long 12.7 mm OD pellets or six 12.7 mm-long pellets. The assembly holds four piezo timing pins. Butted up against the final pellet so that the detonation shock hits it head-on (that is, at normal incidence) is a 0.425 mm-thick Al or 0.254 mm-thick Cu witness disk of slightly more than 19 mm OD. Data consist of PDV traces from a probe with laser beam coincident with axis of symmetry. As for SAX, the PDV probe amply resolves the initial shock jump-off, acceleration, and subsequent multiple shock reverberations within the witness metal (Figure 5).

## 4 Model as Tested: `salinas 1.0`

### 4.1 Functional Form Ansatz

In what follows I now shift to describing the model as I initially conceived, developed, implemented, and tested it.

Following fairly standard practice, I assume that, within any given small but not infinitesimal Lagrangian control volume (*e.g.* cell or zone, in the case of a mesh-based simulation code) of volume $V$, there exist a large but finite number of potential hot-spot initiation sites. I assume that each potential hot-spot initiation site (or just "site") will become an active hot-spot only if hit by a sufficiently strong shock. Subsequent to initiation, within $V$ there exists a fraction $F$ of fully reacted product; the remainder, $(1-F)$, is assumed to be unburned HE. I write, as an ansatz, for the rate[2],

$$\frac{dF}{dt} = G \cdot \mathcal{R}(s) \cdot \mathcal{P}(p) \cdot \mathcal{F}(F), \tag{1}$$

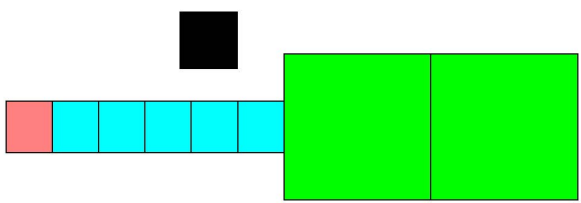

**Figure 3.** ECOT geometry (longitudinal section). Black square is 1 cm × 1 cm. Pink region is booster, light blue region is PBX 9501, and light green is MUT (*e.g.* PBX 9502). Support structure not shown.

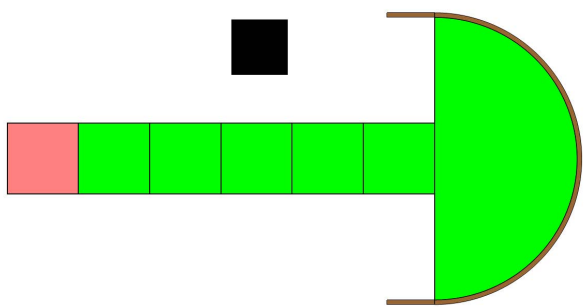

**Figure 4.** SAX geometry (longitudinal section). Black square is 1 cm × 1 cm. Pink region is booster; light green is MUT (*e.g.* PBX 9502), dark brown is metal (Cu or Al) witness shell. Support structure not shown.

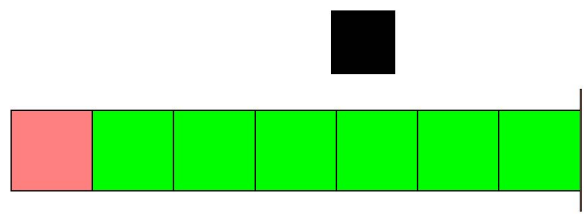

**Figure 5.** DAX geometry (longitudinal section). Black square is 1 cm × 1 cm. Pink region is booster; light green is MUT (*e.g.* PBX 9502). Vertical line at right is metal (Cu or Al) witness disk. Support structure not shown.

---

[2] Note that, formally, $dF/dt$ here indicates the material, or advective, derivative, which would elsewhere often be written as $DF/Dt$ in fluid mechanics.

where $s$ is some thermodynamic quantity as yet to be determined, and $p$ is either the thermodynamic pressure $P$ or, following JT11, the effective pressure $P+Q$ where $Q$ is the isotropic pressure-like stress due to artificial viscosity. That is, the rate is proportional to the product of an initiation function $\mathscr{R}$, a function of pressure $\mathscr{P}$, and a form-factor function $\mathscr{F}$. In my initial work at LLNL, I adopted the functional form of eq. (1) purely by assumption. In developments subsequent to my work at LLNL (`salinas 2.0`, *infra*), I put this assumption on more solid ground. The question of choice for $s$ and $\mathscr{R}$ consume most of my additional work and key assumptions on the initial model (model 1.0); my choices for $\mathscr{P}$ and $\mathscr{F}$ are rather pedestrian by comparison.

Hints for a good choice for $s$ were taken from JT11 and CREST, which have both adopted a similar idea – that is, that initiation in an RFM model should begin with some measure of shock intensity based on a history variable (or a variable that tracks history) – with different answers. Indeed, the model as I initially developed it was heavily influenced by the innovations and successes of JT11 [33]. In CREST, the strength of the shock is determined by the entropy jump [10,11], which, while not formally a history variable, acts like one due to the Second Law. This has a fundamental physical appeal, but entropy is not easy to measure. I therefore chose instead to take the lead from JT11 and relate the strength of a potentially-initiating shock by the history field quantity $p_{max}$, the maximum pressure or effective pressure reached at a given Lagrangian point by the passing shock. Of course, the choice of only a single parameter $s$, rather than multiple, is already a grossly simplifying assumption; see, *e.g.*, higher-order rate improvements to CREST [36] to address just such a concern. Perhaps more importantly, there is empirical evidence to suggest that it may be better to use some combination of $p_{max}$ and the duration of the high-pressure spike [37,38] rather than $p_{max}$ alone. The need for such improvements is not surprising and indeed can be argued on very simple physical grounds. I leave such considerations for future work however. Then, in equation (1) above, I take

$$s \rightarrow p_{max}. \tag{2}$$

This is the second key assumption. The third and final key assumption is described below.

### 4.2 Initiation Function $\mathscr{R}$

Assume that each site within $V$ has a different critical $p_{max}^{(site)}$ drawn from some probability distribution $P(p_{max})$. By *critical pressure*, I mean that a passing shock must reach $p_{max} \geq p_{max}^{(site)}$ in order to turn the site into an active hot spot. I seek the function $P$, and I suggest that the most physically-reasonable form to assume for it, given the limited constraints and information available to the modeller, is that it is log-normal. Solid HE is a complex composite material which, like concrete or asphalt, has a distribution of constituents' sizes – and other properties – spanning decades on a log scale. Various factors may affect a site's shock sensitivity, from its size, to its morphology [39], to its location relative to grain(s), to the grain's crystallinity, to the orientation of exposed crystal surfaces, to the particulars of its amination history. Let us suppose that the various factors that affect the sensitivity of an individual site, as measured by the site's critical $p_{max}$, are all positive-definite and all combine in a multiplicative fashion. At the least, I argue, this is much more physically sensible than assuming they combine additively. The product of a large number of uncorrelated finite random positive-definite factors approaches a log-normal distribution, by the Central Limit Theorem. The log-normal distribution has the nice property, not coincidentally, that it goes to zero at the origin. One can of course imagine any number of counter-arguments, but again, in the absence of more information, I claim that a log-normal form is the best Bayesian prior assumption to adopt for $P$. And so I adopt it. This is my third key assumption. Then

$$P(p_{max})\, dp_{max} = \frac{1}{p_{max}\sigma\sqrt{2\pi}} e^{-(\ln p_{max} - \ln P_\mu)^2 / 2\sigma^2}\, dp_{max} \tag{3}$$

and $P(p_{max}) = P(p_{max}; \sigma, P_\mu)$, where $\sigma$ is a width parameter and $P_\mu$ is the median (Figure 6).

To arrive at $\mathscr{R}$, I simply assume that, immediately following a passing shock of strength $p_{max}$, the burn rate is proportional to the number density of active hot-spots[3], which in turn is proportional to the cumulative distribution function (CDF) (Figure 7) corresponding to the PDF in equation (3). Then

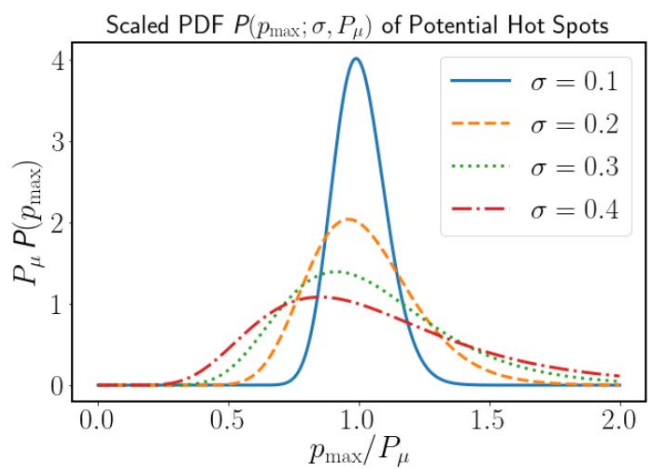

**Figure 6.** PDF of shock-sensitivity of sites.

---

[3] A more careful analysis, see 2.0, shows that, when written in form (1), the rate is proportional to the *cube root* of the number density of active hot-spots.

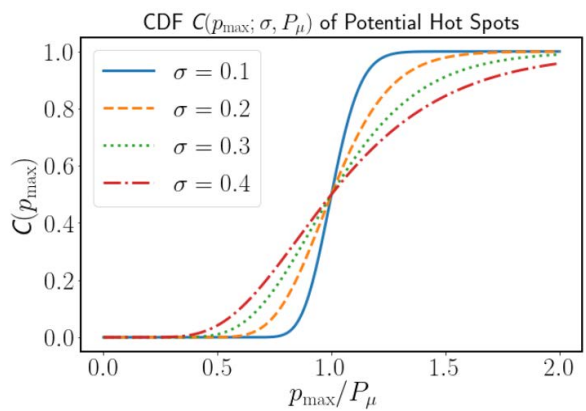

**Figure 7.** CDF of shock-sensitivity of sites.

$$\mathcal{R} = \mathcal{R}(p_{\max}; P_\mu, \sigma) = \Phi\left(\frac{\ln p_{\max} - \ln P_\mu}{\sigma}\right) \quad (4)$$

where $\Phi$ is the normal cumulative distribution function

$$\Phi(x) = \frac{1}{2}\left[1 + \mathrm{erf}\left(\frac{x}{\sqrt{2}}\right)\right], \quad (5)$$

And erf is the error function

$$\mathrm{erf}\, x = \frac{1}{\sqrt{\pi}} \int_{-x}^{x} e^{-y^2}\, dy. \quad (6)$$

### 4.3 Pressure-Scaling Function $\mathcal{P}$

Since, following the initiation of detonation by the von Neumann spike, the subsequent burn in the RZ is a deflagration process on the subgrid (or unresolved) scale, I assume a simple power-law form for the function $\mathcal{P}$, so that $\mathcal{P} \propto p^b$. Physically, I prefer to write $\mathcal{P}$ as a function of the pressure normalized to the C–J pressure; this choice non-dimensionalizes the function $\mathcal{P}$ and gives the rate constant $G$ sensible physical dimensions. Then:

$$\mathcal{P} = (p/P_{\mathrm{CJ}})^b. \quad (7)$$

The factor $P_{\mathrm{CJ}}^{-b}$ can, of course, if one prefers, be absorbed into the rate factor $G$, by defining $G' = G P_{\mathrm{CJ}}^{-b}$ and $\mathcal{P}' = p^b$. Both conventions have their advantages and disadvantages.

### 4.4 Form-Factor Function $\mathcal{F}$

The function $\mathcal{F}$ reflects the form-factor of the rate model, that is, it reflects how burn rate depends upon the fraction $F$ of burned versus unburned HE. After surveying the literature and discussing the matter with various experts, for the `salinas 1.0` rate model, I chose

$$\mathcal{F} = F^{2/3}(1 - F). \quad (8)$$

The exponents on $F$ and $(1 - F)$ were held fixed in all of my work performed at LLNL. The rationale for the 2/3 exponent on $F$ was a simple familiar geometrical argument based on spherical burn fronts propagating outwards from point-like hot-spots. The rationale for the exponent 1 on $(1 - F)$ was simply a phenomenological observation that was relayed to me that, in various models, including the VPLC class of models, this seemed to fit certain CT experiments fairly well.

### 4.5 Equation of State

Following JT11 which was in turn based on JWL++ [13], for the unburned HE (subscript: $u$), I adopt a Murnaghan EOS [40–42], which has two free parameters, $n$ and $\kappa$:

$$P_u = \frac{1}{n\kappa}(v^{-n} - 1) \quad (9)$$

where $\kappa^{-1} = \rho_0 C_0^2$ and $n + 1 = 4S_1$ where $C_0$ and $S_1$ are the shock Hugoniot parameters, and $v$ is the relative volume $v = \rho_0/\rho$.

The burned HE (subscript: $b$) follows a JWL EOS [43–48] in so-called "C-form" (i.e. the adiabat):

$$P_b = Ae^{-R_1 v} + Be^{-R_2 v} + Cv^{-(1+\omega)}. \quad (10)$$

The total thermodynamic pressure in mixed cells is the weighted sum [13], in favour over more computationally-expensive pressure-equilibration in mixed-phase cells (zones):

$$P = (1 - F)P_u + FP_b. \quad (11)$$

This choice sacrifices some physicality in favour of expediency. I was able to justify this based on numerous simulations I performed, examining both the final simulated experimental results as well as the paths in $(v, P)$ thermodynamic state space of fiducial Lagrangian fluid particles. I compared the above EOS based on a weighted sum (11), to a more sophisticated EOS using `CHEETAH 8.0` with proper pressure equilibration in mixed-phase cells (zones). For the immediate task at hand, I did not discern any benefit to the latter approach sufficient to justify its cost.

### 4.6 Complete Model

Combining the above, I now have, for the `salinas 1.0` rate model[4],

$$\frac{dF}{dt} = G\,\Phi\left(\frac{\ln p_{max} - \ln P_\mu}{\sigma}\right)\left(\frac{p}{P_{CJ}}\right)^b F^{2/3}(1-F), \quad (12)$$

where $\Phi$ is as defined in eq. (5), and the overall `salinas 1.0` model is the above rate model combined with the EOS as described above in equations (9–11). (Note that for purposes of actual fitting of simulation to experimental data, I used the effective pressure $P+Q$ for $p$, and I combined $G$ and $P_{CJ}$ into the composite parameter $G' = G P_{CJ}^{-b}$.)

Let me discuss a few properties of this model that I believe recommend it:

- The rate function model in this model separates the shock-initiation process and the subsequent deflagration "burn" process into two distinct functions, $\mathcal{R}$ and $\mathcal{P}$.
- The total number of free parameters for the rate model is small. In all of my work, I left $b$ fixed at the traditional value $b=2$, leaving only three free parameters: $G'$, $\sigma$ and $P_0$. If one were to vary not only $b$ but the exponents on $F$ and $(1-F)$ as well, one would have, in principle, six free parameters, but even this number is small compared with every other model I was aware of and that was otherwise available to me that could fit the behaviour of PBX 9502 in S-COT, ECOT, SAX, and DAX.
- The burn rate $dF/dt$ is a smooth $C^\infty$-function in the interior of the three-dimensional thermodynamic state space $(p_{max}, p, F)$ that is the functional domain of the rate law. There are no abrupt jumps (discontinuities) in $dF/dt$ or any of its derivatives, unlike any other model that was available to me that could fit the corner-turning behaviour of PBX 9502.
- Behaviour $dF/dt$ is a monotonic function of $p$, at fixed $F$ and $p_{max}$. There are no points in the 3-dimensional state space at which rate decreases with pressure.

Regarding the number of free parameters, the modeler of course always feels the need to add additional knobs to turn. Those additional parameters give the modeler freedom. That freedom comes at a definite cost, however. The computational cost alone can quite literally grow exponentially. This means, as a practical matter, that additional resources can not remedy an exploding parameter space. At some point, systematic calibration of a model becomes not just expensive, but physically impossible. For this reason among others, I view it as a noteworthy goal to develop a model that, on the one hand, can capture the salient features of the phenomena of interest (*e.g.* corner turning) while, on the other hand, do so with a relatively small number of free parameters.

### 4.7 Implementation

I implemented and tested the above model in an ALE code at LLNL that is widely-used for reactive-flow simulations, dynamic simulations of HE and metal, and other applications. For expediency, I simply created my branch of the main LLNL ALE code base and re-wrote the C-language function call to the JT11 rate law to use the `salinas 1.0` rate law as described above.

The LLNL ALE code that I used (hereafter: ARES) is a massively-parallel, multi-block ALE-AMR (Arbitrary Lagrangian-Eulerian-Adaptive Mesh Refinement) multi-physics code, written in C, with a 20-year pedigree at LLNL. Applications include HE, laser-driven implosion, and magnetically-driven experiments (Figure 8).

The AMR aspect consists of refinement restricted to have neighboring cells (zones) differ by at most one level of refinement. Refinement is hierarchical with a single global time step. Levels of refinement can have variable refinement ratios constrained to be odd numbers, *e.g.* 3×3 or 5×5 for a 2D problem. The maximum number of levels of refinement is for all practical purposes limited by computational resources rather than the code itself. The AMR includes refinement based on various physics or numerical flags or conditions and also includes de-refinement that can be made suitably aggressive to avoid over-refinement in regions long after the passage of a strong, thin shock. Refinement criteria may include such quantities as second differences, first differences, field values, and mesh distortion, as well as other user-defined criteria.

Hydrodynamics consist of single-fluid multi-material staggered-mesh hydrodynamics. Timesteps consist of a purely Lagrangian step plus, optionally, an approximately 2$^{nd}$-order accurate monotonic flux-based remap step with equipotential mesh relaxation. In contrast with the relatively better-known LLNL code ALE-3D, with which ARES shares much in common (but also much *not* in common), ARES takes explicit, rather than implicit, time-steps. The La-grange step is, essentially, a 2nd-order accurate predictor-corrector. The hydrodynamics is Navier-Stokes, with the option to use either analytic or tabular viscosity.

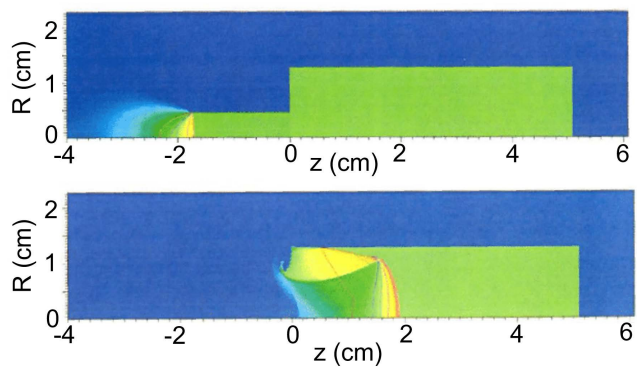

**Figure 8.** Two instants in time of one of my simulations of corner-turning in ECOT. Color indicates density; light green is nominal PBX 9502 density of 1.89 g/cc.

---
[4] There is an additional small kick or offset in burn fraction to initiate the burn.

The problem generation workflow includes the writing, by the user, of an ASCII input deck that is processed by an internal parser with simple control-flow constructs. The user has the option to augment his or her input deck with blocks of either Python or Lua included within the input deck.

Simple geometry mesh specification is typically done in-line, meaning, following explicit mesh-generation commands written by hand in the input deck. More complex geometries, as I learned after the fact, can be created with a variety of LLNL tools, including PMESH. PMESH particularly simplifies the construction of multi-block problems in which logically-structured blocks are sewn together in a manner that may be globally, to a degree, unstructured. Restrictions on global mesh topology however are not trivial, as the meshes of sewn-together blocks must match along their shared mutual edges (or surfaces). Output results can be processed with various visualization tools and transferred to other meshes (such as to generate synthetic X-ray radiographs) via the LLNL code Overlink.

ARES has been successfully applied to study instability growth in laser-driven implosion mix experiments, as well as resultant DT/TT fusion burn. It also possesses various plasma physics, MHD, and radiation transport physics, but these are of no relevance to the present work. Material models include a range of EOS options including analytic and tabular forms, as well as material strength and failure models such as Steinberg-Guinan strength model [49], which I used for witness metal in SAX and DAX, among other models. Finally, ARES is able to treat material interfaces with so-called slide surfaces. This method is particularly useful for problems in which HE abuts metal, such as in the SAX and DAX problems.

Although in principle ARES output can be visualised by other means, typically the tool of choice is the LLNL code VisIt, and this was the workflow followed here. In addition, of particular note for the present work, ARES has the capability to simulate virtual laser PDV probes, and can, therefore, be applied to the simulation of both SAX and DAX, which have metal witness surfaces probed by laser PDV (Figure 9).

For ECOT, SAX, and DAX, my simulations were as follows: initial resolution of 30 cells per cm, with a maximum of three (3) levels of mesh refinement. Each AMR refinement level replaced a single cell (zone) with nine cells (zones) in a 3x3 pattern, with maximum resolution limited to 400 cells (zones) per cm. (Note that, being an ALE code, the size of the coarsest level of meshing does not remain at 30 cells per cm, and so in principle, 3 levels of AMR refinement are possible despite the restriction of a maximum of 400 zones per cm.) Initial time step was taken to be $10^{-4}$ μs, but this time step was dynamically reduced during the simulation to accommodate high-resolution AMR. The mesh was relaxed to avoid tangling, via modified equipotential mesh relaxation. The EOS parameters (JWL, Murnaghan) varied, but typical fiducial JWL parameters for PBX 9502 were: $R_1 = 4.60$, $R_2 = 1.51$, $A = 6.74$, $B = 0.157$, $\omega = 0.24$, $\Gamma + 1 = 3.93$, and $E_0 = 0.076$ Mbar-cc/cc (these latter values informing $C$ for the JWL adiabat)[5]. These parameters were written in the input deck with a custom syntax, allowing a master Perl script to search-and-replace these values on the fly. This allowed the entire simulated experiment to be rapidly repeated with a different set of input values for the EOS and the `salinas 1.0` rate law, and results evaluated. EOS and strength parameters for the witness metal in DAX and SAX were taken from standard tables available at LLNL; these specific values do not materially inform the discussion here, however.

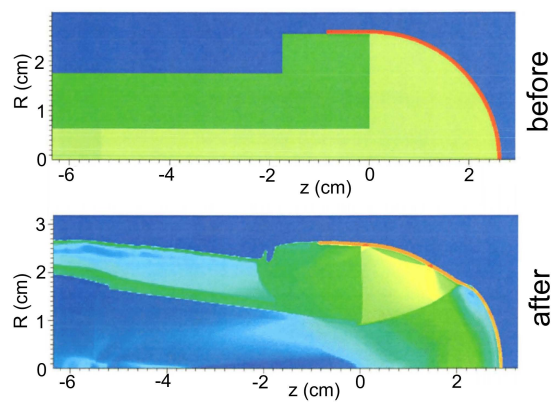

**Figure 9.** One of my simulations of SAX with ARES, showing initial configuration and intermediate time. Shade indicates density. Medium green is plastic holder; light green is HE. Witness shell appears ruddy brown in initial configuration. Range of densities is automatically re-scaled, and hence witness does not appear ruddy brown in lower instant in time.

### 4.8 Preliminary Results

Beginning in mid-2016 I performed initial testing using my input deck for S-COT and found quite satisfactory results. Recall S-COT data includes a streak-camera record of the shock breakout from the side of the acceptor cylinder. In simulating this, my model showed run speeds comparable to JT11 and an approximate 20x speedup over VPLC runtimes. This speedup did not come at the cost of any loss of fidelity. Rather, with a physically-reasonable choice of rate parameters for my model (*infra*), combined with EOS parameters taken from previous calibrated tests, I was able to fit S-COT data quite well. Qualitatively, the fit was as good as any fit that I have witnessed or seen published by either

---

[5] Note that Mbar is an energy density; the somewhat idiosyncratic notation here is used, at LLNL at least, to indicate a Lagrangian, rather than Eulerian, energy density

VPLC or JT11 models. Unfortunately, however, again, I no longer have access to any of that test data or the specific values of the rate law parameters that were seen to fit S-COT data.

Serious application of `salinas 1.0` to ECOT, SAX, and DAX, was not undertaken until early 2017 after I wrote input decks to run the ALE code for each of these experiments. Ideally, subsequent fitting would involve a rigorous multi-dimensional maximization procedure involving some kind of objective fitness function. For expediency, however, I performed simple parameter sweeps, and, following common practice, I simply judged fit by eye. (This appears to be the standard in the field, in part, one assumes, because it is more difficult to construct an objective fitness function than one might think.)

For the immediate purposes at hand, it was suggested to me that I needed to adjust the EOS of unburned HE to capture lot-to-lot variations[6], including the EOS of the unreacted HE. The full parameter set that I considered then was the following: Murnaghan $n$ and $\kappa$ for the unburnt HE, and $G'$, $P_0$, and $\sigma$ for the rate model. In all cases, for any one set of parameter values, I insisted not in judging goodness of fit for only one experiment, but all three simultaneously. By this process, I was able to obtain simulated results for ECOT, SAX, and DAX that were, at least qualitatively, the equal of any other fit that I witnessed for any model, either VPLC or JT11 when applied to any more than one single choice experiment. I viewed these results as encouraging.

## 5 Improved Model: `salinas 2.0`

### 5.1 Formal Development

Subsequent to my initial model development, implementation, and testing at LLNL as described above, I have attempted to put the model on more solid theoretical footing, with some modifications as described below. I describe this improved model separately, as it has not yet been tested.

As before, let us assume that potential hot-spots are identical in all aspects except for one: that each has a critical pressure $p_{max}^{(site)}$ drawn from some probability distribution $P(p_{max})$. That is, for any given potential initiation site, the probability of that site's critical $p_{max}^{(site)}$ being between a given $p_{max}$ and $p_{max} + dp_{max}$ is simply $P(p_{max}) dp_{max}$. The corresponding cumulative distribution function (CDF) is

$$C(p_{max}) = \int_0^{p_{max}} P(p'_{max}) dp'_{max}, \qquad (13)$$

and $\lim_{p_{max} \to \infty} C(p_{max}) = 1$.

Within the small but not infinitesimal test region of volume $V$, let $\chi$ represent the number density of potential hot-spot initiation sites, and as before, let $1 \ll \chi V$. Immediately after passage of a strong shock that has reached a maximum pressure of some $p_{max}$, let $\chi_*$ denote the number density of actual (active) hot-spots. Then $\chi_* = \chi C(p_{max})$. Within $V$, on the subgrid-scale, let there be a spherical deflagration front burning outward from each active hot-spot. These burn-fronts will of course merge over time. Suppose that, at any given instant in time, the speed of this deflagration front or fronts is the same everywhere within $V$, independent of hot-spot, but dependent upon pressure $p$. Let us write the speed (as defined in Lagrangian coordinates) as

$$u_b = u_{bCJ} \mathscr{P}(p), \quad \mathscr{P}(P_{CJ}) = 1, \quad \mathscr{P}(p) = (p/P_{CJ})^b. \qquad (14)$$

The test volume will undergo change in (Eulerian) volume, not to mention other deformations; the question arises how to define the subgrid-scale burn front speed $u_b$ in the face of these strains. My choice, again, is to define the burn front speed in Lagrangian coordinates, and to neglect, in this derivation, all strains other than uniform compression or dilatation. This is in keeping with the highly-simplifying assumptions adopted throughout this model. The burn front speed in Eulerian co-ordinates is then $u_b v^{1/3}$ where $v = \rho_0/\rho$ is the volume expansion factor, but this Eulerian speed is not required of the model.

An expression for the overall burn rate may be derived beginning with the following familiar argument. Immediately after shock initiation leading to a number density $\chi_*$ of active hot-spots, the number of actual discrete active hot-spot initiation sites within the finite test volume $V$ follows a Poisson distribution with mean $\chi_* V$. The probability of finding zero sizes is then $\exp(-\chi_* V)$, and the probability of finding at least one site is of course $1 - \exp(-\chi_* V)$. Given a trial probe location positioned at random then, we write $P_{dnh}(r) dr$ to indicate the probability that the distance $r'$ from the trial probe to the nearest hot-spot is between $r$ and $r + dr$. This probability is simply the product of the probability of finding zero sites inside the volume $4\pi r^3/3$ within $r$ of our probe, times the probability of finding at least one site inside the volume $4\pi r^2 dr$ between distance $r$ and $r + dr$ of our probe.

An expression for $P_{dnh}(r)$ may be found by taking the limit that this latter probability is small so that $(1 - e^{-x}) \simeq x$. As $F$ represents the fraction of HE that has burned as a function of time $t$, then $dF/dt = u P_{dnh}(r)$ where now $r = \int^t u_b \, dt$, and we have

$$\frac{dF}{dt} = 4\pi \chi * u_b r^2 e^{-4\pi \chi_* r^3 / 3}. \qquad (15)$$

---

[6] One might argue that the EOS should not vary from lot to lot. I do not argue this point one way or another.

This may be integrated to find the familiar [50]

$$F = 1 - e^{-4\pi\chi_\star r^3/3}, \tag{16}$$

a result that could have been easily anticipated. Combining the two equations above, we have, trivially, and recalling $\chi_\star = \chi C(p_{max})$ and my expression (14) for $u_b$,

$$\frac{dF}{dt} = G\,C^{1/3}(p_{max})\,\mathcal{P}(p)\,[-\ln(1-F)]^{2/3}(1-F) \tag{17}$$

Where

$$G = (36\pi)^{1/3}\chi^{1/3}u_{bCJ} \simeq$$
$$48.4\mu s^{-1}\left(\frac{\chi}{10^3\,\text{mm}^{-3}}\right)^{1/3}\left(\frac{u_{bCJ}}{1\,\text{mm}\,\mu s^{-1}}\right), \tag{18}$$

and $G$ has the units of a rate, $[1/T]$. I can now identify the form-factor function $\mathcal{F}(F)$ as

$$\mathcal{F}(F) = [-\ln(1-F)]^{2/3}(1-F), \tag{19}$$

and the initiation function $\mathcal{R}(p_{max})$ as

$$\mathcal{R}(p_{max};\sigma,P_\mu) = C^{1/3}(p_{max};\sigma,P_\mu) =$$
$$\Phi^{1/3}\left(\frac{\ln p_{max} - \ln P_\mu}{\sigma}\right). \tag{20}$$

Figures 10–13 illustrate the reasoning and phenomena as described above that led to this expression.

### 5.2 Initial Discussion

This result for $\mathcal{R}$, eqn. (20), satisfies a few pleasing properties[7]:
1. $\mathcal{R}(p_{max}) \to 0$ as $p_{max} \to 0^+$. An infinitesimally-weak shock will not initiate burn anywhere.
2. $\mathcal{R}(p_{max}) \to 1$ as $p_{max} \to \infty$. A sufficiently strong shock will initiate burn at all potential hot-spots.
3. All derivatives of $\mathcal{R} \to 0$ both as $p_{max} \to 0^+$ and $p_{max} \to \infty$. In particular, regardless of the (finite, positive) value of $P_\mu$ and $\sigma$, the initiation curve is not just zero but *flat* with respect to shock strength for infinitesimally-weak shocks.
4. The width parameter $\sigma$ can take any positive value; for example, properties (1–4) above still formally hold for arbitrarily large (but finite) values of $\sigma$.
5. The function $\mathscr{R}$ is a monotonic $C^\infty$-function with a sigmoid shape. Experience has shown that TATB appears to be best fit with a sigmoid curve (with respect to pressure or shock strength) appearing somewhere in the rate model, and here we have a function with such a shape.
6. Unlike other candidate ad-hoc sigmoid functions, such as tanh, the log-normal function is physically-motivated, being based on the simple model and derivation given above.
7. The function $\mathscr{R}$ has only two free parameters, $P_\mu$ and $\sigma$.

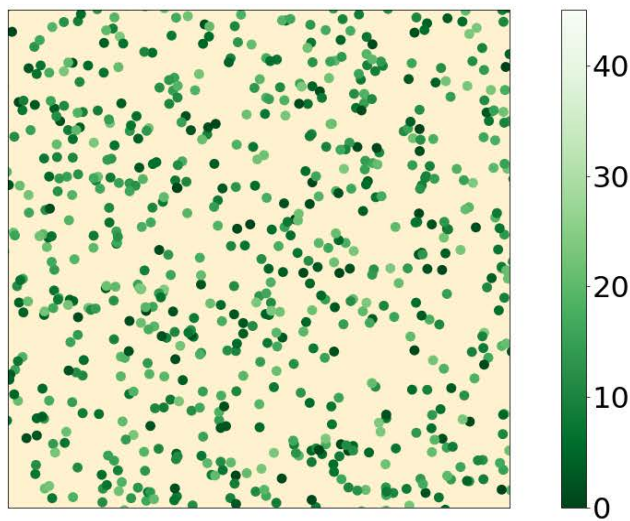

**Figure 10.** Sites have a range of sensitivities. Colour bar is sensitivity in GPa.

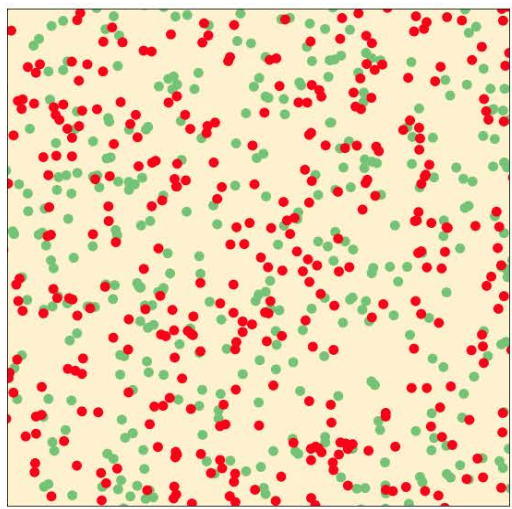

**Figure 11.** After a passing shock, only some sites become active (dark red).

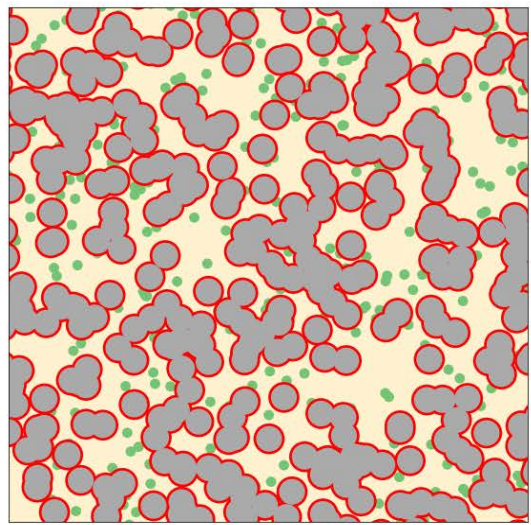

**Figure 12.** Burn proceeds in a thin layer (dark grey) of conjoined spherical shells.

---

[7]These properties were also satisfied by the earlier 1.0 version of $\mathcal{R}$, eq. (4).

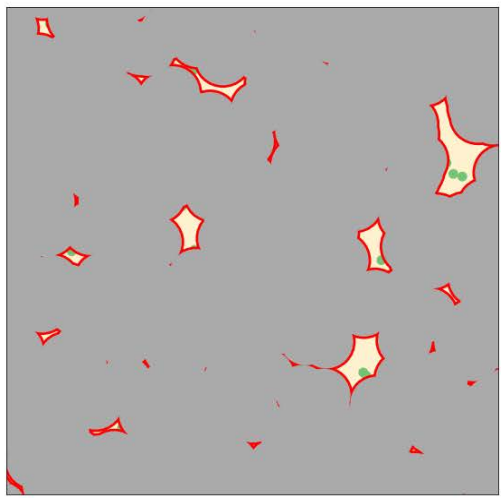

**Figure 13.** Form-factor function reflects resultant total surface area of burn front as a function of burn fraction. Here we see the nearly-final stages of burn.

Regarding (5) and (6), previous work, including much work by others, such as but not limited to the JT11 model and the VPLC model, as well as other unpublished work by the author done in collaboration with P. Vitello and K. T. Lorenz at LLNL, seem to indicate, again, that TATB is best modeled with a sigmoid-shape pressure sensitivity. Here, that sigmoid shape is encapsulated in the initiation function $\mathscr{R}$, leaving a nice and simpler power-law dependence for the post-initiation burn as described by the pressure modulation function $\mathscr{P}$.

Combining all the above, I re-write equation (17) as

$$\frac{dF}{dt} = G \cdot \mathscr{R}(p_{\max}) \cdot \mathscr{P}(p) \cdot \mathscr{F}(F), \qquad (21)$$

recovering what initially began as an ansatz, equation (1), in the earlier version of my model. Based on my experience with a previous as-yet-unpublished VPLC-based model of mine, the so-called "bluff" model, as well as my recollection, representative good values for rate parameters for PBX 9502 are very approximately the following: $G = 50\,\mu s^{-1}$, $P_\mu = 20$ GPa, and $\sigma = 5$ GPa.

### 5.3 The Model Proper

To summarize, `salinas 2.0` consists of the rate law (21), initiation function (20), burn speed scaling function (7), and form-factor function (19), combined with the JWL++-style EOS as given in (9-11); the rate constant may also be related to site number density and deflagration front speed as in (18). The pressure-modulation function, as well as the EOS, are unchanged from `salinas 1.0`, and the expression for the rate constant $G$ simply relates this quantity to the site number density $\chi$ and the deflagration front speed $u_{bCJ}$ at the CJ pressure. From a fitting perspective then, the only substantial changes are the $1/3$ exponent in the relation for $\mathscr{R}$, and the new form for $\mathscr{F}$. I point this out because while I was able to test `salinas 1.0` fairly extensively on S-COT, ECOT, SAX, and DAX, I have not tested `salinas 2.0`. It is therefore worth discussing whether these improvements to `salinas` affect the ability of the model to be a suitable choice for CT and other HE experiments that one might wish to model with RFM.

### 5.4 Anticipated Effects of Improvements

I argue that the addition of the $1/3$ exponent on the log-normal CDF $\Phi$ in going from `salinas 1.0` to `salinas 2.0` makes a negligible difference, practically speaking, resulting mainly in a shift in the fitted values of $\sigma$ and $P_\mu$. For example, the function $\Phi^{1/3}(p_{\max}; \sigma, P_\mu)$ with $\sigma = 0.1$ and $P_\mu = 1.0$ is hardly visually distinguishable from the function $\Phi(p_{\max}; \sigma, P_\mu)$ with $\sigma = 0.14$ and $P_\mu = 0.89$. The same is true for $\Phi^{1/3}(p_{\max}; \sigma, P_\mu)$ with $\sigma = 0.3$ and $P_\mu = 1.0$ compared with $\Phi(p_{\max}; \sigma, P_\mu)$ with $\sigma = 0.42$ and $P_\mu = 0.7$ (see Figure 14).

The change in the functional form for the form-factor function $\mathscr{F}$ is more substantial, but not, I believe, so substantial so as to cause `salinas 2.0` to fail as a model where `salinas 1.0` did not. In other words, I expect that `salinas 2.0` will prove to be just as suitable for simulating S-COT, ECOT, DAX, and SAX as `salinas 1.0` has already been found to be. The greatest difference is a relative speed-up in the completion of the burn for 2.0 when compared to 1.0 (see Figure 15).

### 5.5 Discussion and Future Plans

In `salinas 2.0`, just as with `salinas 1.0`, I note:

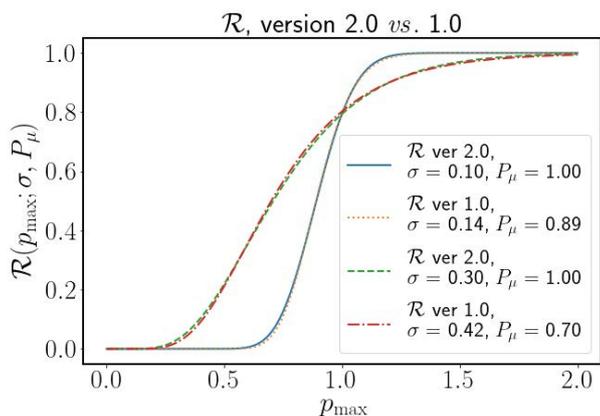

**Figure 14.** As a practical matter, the difference between ver 2.0 and 1.0 of the initiation function is largely a matter of a shift of σ and $P_\mu$.

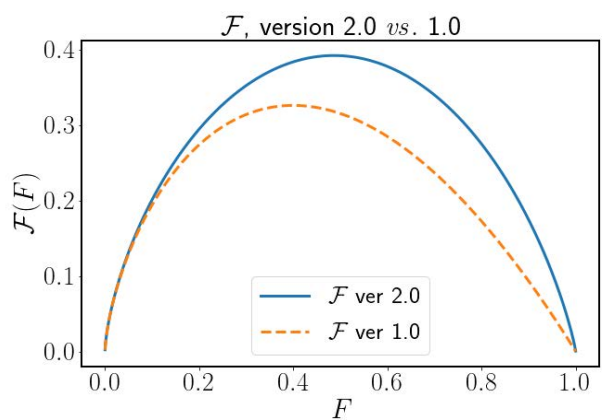

**Figure 15.** Illustration of the change in the form-factor function in going to version 2.0 from version 1.0 of the model.

- The rate function model separates the initiation process and the subsequent burn process into two distinct functions, $\mathcal{R}$ and $\mathcal{P}$.
- The total number of free parameters for the rate model is small, as few as three or four in practice.
- The burn rate $dF/dt$ is a smooth $C^\infty$-function in the interior of the functional domain of the rate law. The model eschews the step-function jump in $dF/dt$ JT11 that I believe led to stability problems.
- Unlike the VPLC class of models discussed above, at fixed $F$ and $p_{max}$, the behavior $dF/dt$ is a monotonic function of $p$.
- The model is, judging from my experience with `sali-nas 1.0`, most likely about 20x as fast as 2017-era VPLC models that used `CHEETAH 8.0`.

The smoothness of the rate law is shown in Figure 16. Here we see iso-surfaces for the rate $dF/dt$ in the three-dimensional thermodynamic state space $(p_{max}, p, F)$. Darker colours correspond to higher rates. Note that the rate law is only defined for $p \leq p_{max}$, which, within the figure, is the half of the cube closer to the reader. Also shown is a notional path of a Lagrangian fluid particle in this state space, starting from the lower-left corner. The path in 3D is also projected onto the three 2D planes, as shown by the three-dotted lines. I have found that there is much value in examining the actual simulated paths in the thermodynamic state space for trial Lagrangian test particles, both for this model and for VPLC and JT11.

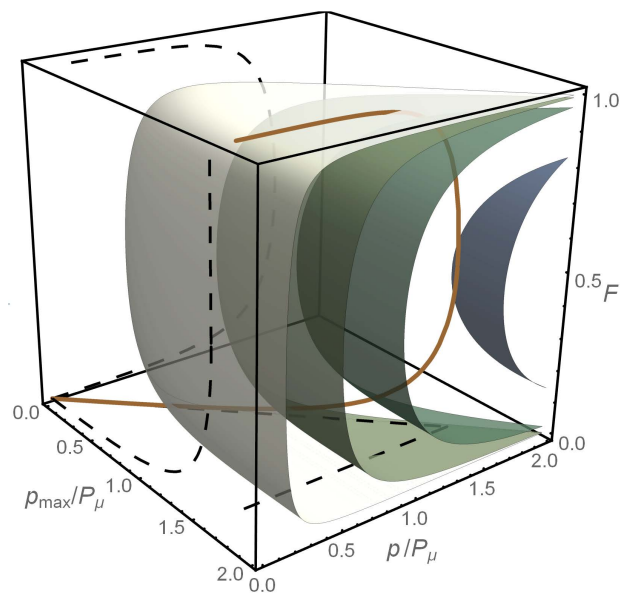

**Figure 16.** Iso-surfaces of rate $dF/dt$ (darker colours are higher rates), and notional path of Lagrangian fluid particle in thermodynamic state space. See text for further explanation.

In Figure 17 one sees the time history, following the same Lagrangian fluid particle as shown in Figure 16, of the state variables $p_{max}$, $p$, and $F$, as well as the rate $dF/dt$. Being entirely notional, the scale of the time axis has no particular significance here, and so $dF/dt$ is re-normalized to have a maximum value of 1.0 so that it may easily be viewed on the same graph. I can and did perform this exercise with actual simulation data of a suite of Lagrangian test particles, both for `salinas 1.0` and for VPLC and JT11 models, and found this exercise quite informative.

There are a number of future modifications to this model one can imagine, the most important of which involves additions to accommodate a range of densities. It is well-known that corner-turning in PBX 9502 and LX-17 is ex-quisitely sensitive to the density to which the HE has been pressed, an effect that has important real-world consequences. For the same lot, an over-dense pressing results in relatively poor corner-turning, whereas an under-dense pressing results in more abrupt corner-turning. To accommodate this behavior in VPLC models, one has to go in by hand, introducing a fudge factor [14] to turn up the rates for over-dense pressings (and turn down the rates for under-dense pressings); without this, models such as the VPLC model will predict behavior that is the exact opposite of what is seen on the test stand. A similar approach might be needed here to adjust the rate model to accommodate a range of densities above and below nominal. The problem is, one is not free to do this: as mentioned above, blindly increasing the rate to better-match ECOT for a given lot worsens the fit for DAX. This is as true for VPLC and JT11 as it is for my model. This is one of many instances where a reduced dimensionality of the rate-law parameter space comes into play, because without this, the global minimization problem of finding a suitable rate law to accommodate lot-to-lot variations in all three experiments, rather than just one, is prohibitive.

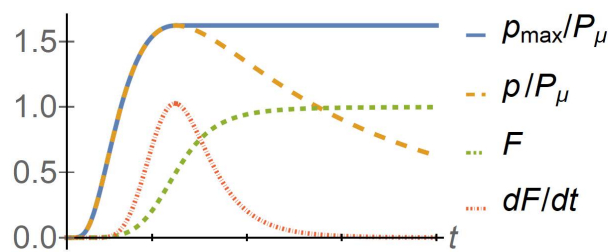

**Figure 17.** Notional behavior versus time of thermodynamic state and reaction rate. For illustration purposes, rate is re-normalized here so that max(rate) = 1.0.

Additionally, one might anticipate that, to the extent that small voids created in the pressing process may contribute to the budget of potential hot-spots, $\chi$, and hence $G$, and $P_\mu$ as well, might be dependent upon HE density. Similarly, I can anticipate modifications to the EOS as well, the most important of which is probably the inclusion of the HE void fraction as represented by the complement of the %TMD to which the HE is pressed. A sensible approach is a snowplough-type or crush-rate [51] modification the EOS. It would be hoped that such additions, both to the rate law and the EOS, would enable the overall model to be a suitable choice for capturing the effect of density on corner-turning performance.

Absent such density-aware modifications however, I consider the EOS for `salinas 2.0` to be unchanged from `salinas 1.0`: a two-parameter Murnaghan EOS for unreacted HE, a C-form JWL for reacted HE, and a simple linear mixer for mixed-phase cells.

## 6 Conclusion

My initial development of `salinas 1.0` began because I was asked to develop a model to capture CT in new lots of PBX 9502. It eventually became clear to me that what was meant by this was not to construct a new model, but rather to find suitable parameter values for already-constructed models; at this point, at or around the time I found `salinas 1.0` model to fit S-COT well, my work on `salinas` became a side-project. This proved quite fortuitous later however when I encountered insurmountable obstacles reproducing experimental results with the other models available to me.

Among other difficulties, as I have noted, VPLC models are computationally intensive and have a large number of free parameters, which can be a challenging combination. In addition, as I discovered, while `CHEETAH 8.0` as used in VPLC models may be relatively sophisticated as far as its EOS for fully-reacted HE and its method for finding the EOS of mixed cells (zones) may be concerned, it employs a two-parameter Murnaghan EOS for the unburnt HE. This would not be such a problem but for another fact: the Murnaghan parameters $n$ and $\kappa$ for unburnt HE are specified in a binary file; they are not modifiable nor even knowable to the casual user. As I required the ability to adjust the EOS of the unburnt HE, this ruled out VPLC models.

Regarding the JT11 model, I note that I encountered what appeared to be unstable behavior in the lightly-confined rate stick donor for DAX and SAX. This was first discovered upon examining convergence studies that I was instructed to perform. These studies showed that the jump-off speed of DAX followed a bimodal distribution in simulations. Upon further investigation, I discovered that about half the time, the HE immediately adjacent to the metal witness disk did not fully burn. This can be a subtle effect if one looks only at pressure or density, but it stands out dramatically if, as I did, one visualizes burn fraction $F$ with a suitable tool such as VisIt (my visualization tool of choice).

The `salinas 1.0` model did not appear to suffer from this problem, and I found no example running S-COT, SAX, DAX, or ECOT at any resolution where `salinas 1.0` did not perform as well or better than JT11. This is true both speaking generally, as well as specifically regarding ability to fit experimental data.

Unfortunately, I am no longer in possession of any of my simulation data demonstrating the performance of `salinas 1.0`, and I of course have no data regarding `salinas 2.0`, but I hope that other researchers may nonetheless find these models and their development useful or informative. I believe the models may be nicely suited to perform the lot-to-lot calibration of PBX 9502 mentioned at the outset.

Indeed, one of the ultimate goals of this work is to relate performance of designs to known metrics for the HE and its constituents. For example, a key metric is the size distribution of the TATB particles; it is known that this size-distribution is related to performance [52,53]. Ideally, PBX 9502 has a PSD (Particle Size Distribution) largely peaking in the range of 50–100 μm. Assuming a nominal burn-front speed $u_{bCJ}$ of 1 mm$\mu$s$^{-1}$ at the C–J condition, and taking the density of potential hot-spot sizes as one per representative 50–100 μm grain, I obtain a rate parameter $G$ from equation (18) that turns out to be of order of the result I measured from fitting the model to data, where again, very approximately, $G = 50 \mu s^{-1}$, $P_\mu = 20$ GPa, and $\sigma = 5$ GPa. I find this result encouraging.

## Symbols

| | |
|---|---|
| $A$ | JWL EOS high-pressure coefficient |
| $B$ | JWL EOS medium-pressure coefficient |
| $b$ | *subscript*: burned<br>*superscript*: exponent in burn speed pressure-scaling |
| $C$ | JWL EOS low-pressure coefficient |
| $\mathbf{C}$ | CDF for sensitivity of potential hot-spots |
| $C_0$ | shock Hugoniot parameter |
| $C^\infty$ | infinitely differentiable |
| $D_1$ | diameter of first (smaller) cylinder in double-cylinder experiment |
| $D_2$ | diameter of second (larger) cylinder in double-cylinder experiment |
| $E_0$ | JWL parameter |
| $F$ | fraction of HE that has been reacted to product; $0 \leq F \leq 1$ |
| $\mathscr{F}$ | form factor function |
| $G$ | rate constant |
| $G'$ | rate constant (alternative normalization) |
| $n$ | parameter in Murnaghan EOS for unreacted HE |
| $P$ | thermodynamic pressure |
| $P_\mu$ | median potential hot-spot shock sensitivity |
| $P_b$ | pressure of burned HE |
| $P_{CJ}$ | Chapman-Jouguet pressure |
| $P_u$ | pressure of unburned HE |
| $\mathbf{P}_{dnh}$ | PDF for the distance, from any given point to the nearest active hot-spot |
| $\mathscr{P}$ | burn speed pressure-scaling function |
| $\mathscr{P}'$ | burn speed pressure-scaling function (alternative normalisation) |
| $\mathbf{P}$ | PDF for sensitivity of potential hot-spots |
| $p$ | generalized pressure; typically $p = P + Q$ |
| $p_{max}$ | maximum historical value of $p$ at a given Lagrangian point |
| $p_{max}^{(site)}$ | critical pressure to initiate burn at a given potential hot-spot site. |
| $Q$ | pressure-like stress due to artificial viscosity |
| $\mathscr{R}$ | initiation factor function |
| $R_1$ | JWL EOS parameter |
| $R_2$ | JWL EOS parameter |
| $r$ | radius of sphere of burned HE surrounding an active hot spot |
| $S_1$ | shock Hugoniot parameter |
| $s$ | generalized quantity related to initiation effectiveness of shock |
| $t$ | time |
| $u$ | *subscript*: unburned |
| $u_b$ | speed (Lagrangian) of deflagration front spreading out from active hot-spot. |
| $u_{bCJ}$ | speed (Lagrangian) of hot-spot deflagration front at CJ pressure |
| $V$ | volume (of test region) |
| $v$ | relative volume; $v = \rho_0/\rho$ |
| $x, y$ | notional variables |
| $\Gamma$ | JWL parameter |
| $\kappa$ | Murnaghan EOS parameter |
| $\rho$ | density |
| $\rho_0$ | initial density of HE (i.e. density at STP) |
| $\sigma$ | width of distribution of shock-sensitivities of potential hot-spots |
| $\Phi$ | normal CDF |
| $\chi$ | number density (Lagrangian) of potential hot-spot sites |
| $\chi_\star$ | number density (Lagrangian) of actual hot-spots immediately after shock passage |
| $\omega$ | ideal gas exponent, JWL EOS |

*Acknowledgement*

The author thanks numerous former co-workers of his at LLNL, including K. T. Lorenz and many others, for their ongoing encouragement and support. This research was funded in part by the DOE/NNSA under contract DE-AC52-07NA27344.

## References

[1] D. L. Chapman, Detonation Waves VI. On the rate of explosion in gases, *Philos. Mag. (1798–1977)*, **1899**, *47*, 90. doi: 10.1080/14786449908621243.

[2] E. Jouguet, Sur la propagation des réactions chimiques dans les gaz, *J. Math. Pures Appl* **1906**, *2*, 5.

[3] M. E. Berger, *Detonation of High Explosives in Lagrangian Hydrodynamic Codes Using the Programmed Burn Technique*, Report LA-6097-MS, Los Alamos Scientific Laboratory, Los Alamos, NM, USA **1975**. doi: 10.2172/4148515.

[4] J. B. Bdzil, W. C. Davis, *Time-Dependent Detonations*, Report LA-5926-MS, Los Alamos Scientific Laboratory, Los Alamos, NM, USA **1975**.

[5] C. L. Mader, *Two-Dimensional Flow of Detonation Proceeding Perpendicular to Confined and Unconfined Surfaces*, Report LA-UR-73-91, Los Alamos Scientific Laboratory, Los Alamos, NM, USA **1973**.

[6] V. S. Trofimov, A. N. Dremin, Structure of the nonideal detonation front in solid explosives, *Combust. Explos. Shock Waves (Engl. Transl.)*, **1971**, *7*, 368. doi: 10.1007/BF00742826.

[7] J. B. Bdzil, Steady-state two-dimensional detonation, *J. Fluid Mech.*, **1981**, *108*, 195. doi: 10.1017/S0022112081002085.

[8] D. S. Stewart, J. B. Bdzil, The shock dynamics of stable multi-dimensional detonation, *Combust. Flame*, **1988**, *72*, 311. doi: 10.1016/0010-2180(88)90130-7.

[9] E. L. Lee, C. M. Tarver, Phenomenological model of shock initiation in heterogeneous explosives, *Phys. Fluids (1958-1988)*, **1980**, *23*, 2362. doi: 10.1063/1.862940.

[10] C. A. Handley, The CREST Reactive Burn Model, *15th Conference of the American Physical Society Topical Group on Shock Compression of Condensed Matter*, Waikoloa, Hawaii (USA), June 24–29, **2007**, AIP Conference proceedings 955, p. 373. doi: 10.1063/1.2833061.

[11] C. Handley, N. Whitworth, H. James, B. Lambourn, M.-A. Maheswaran, The CREST reactive-burn model for explosives, *EPJ Web Conf.* **2010**, *10*, 00004. doi: 10.1051/epjconf/20101000004.

[12] R. Menikoff, M. S. Shaw, Reactive burn models and ignition & growth concept, *EPJ Web Conf.* **2010**, *10*, 00003. doi: 10.1051/epjconf/20101000003.

[13] P. C. Souers, S. Anderson, J. Mercer, E. McGuire, P. Vitello, JWL++: A Simple Reactive Flow Code Package for Detonation, *Propellants Explos. Pyrotech.* **2000**, *25*, 54.

[14] I.-F. W. Kuo, P. Vitello, L. E. Fried, E. V. Bukovsky, K. T. Lorenz, Reactive Flow Modelling of Small Scale Corner Turning Experiments, *16th International Detonation Symposium*, July 15–20, **2018**, Cambridge, MD, USA.

[15] M. Cox, A. W. Campbell, Corner-Turning in TATB, *7th International Detonation Symposium*, June 16–19, **1981**, Annapolis, MD, USA.

[16] M. Held, Corner-Turning Distance and Retonation Radius, *Propellants Explos. Pyrotech.* **1989**, *14*, 153. doi: 10.1002/prep.19890140406.

[17] M. Held, Corner Turning Research Test, *Propellants Explos. Pyrotech.* **1996**, *21*, 177. doi: 10.1002/prep.19960210403.

[18] C. D. Hutchinson, G. C. W. Foan, H. R. Lawn, A. G. Jones, Initiation and Detonation Properties of the Insensitive Highe Explosive TATB/Kel-F 800 95/5, *Ninth Symposium (International) on Detonation*, August 28 – September 1, **1989**, Portland, Oregon, USA, 123.

[19] P. C. Souers, H. G. Andreski, C. F. Cook, III, R. Garza, R. Pastrone, D. Phillips, F. Roeske, P. Vitello, J. D. Molitoris, LX-17 Corner-Turning, *Propellants Explos. Pyrotech.* **2004** *29*, 359–367, doi: 10.1002/prep.200400067.

[20] P. C. Souers, H. G. Andreski, J. Batteux, B. Bratton, C. Cabacungan, C. F. Cook, III, S. Fletcher, R. Garza, D. Grimsley, J. Handly, A. Hernandez, A. McMaster, P. McMaster, J. D. Molitoris, R. Palmer, J. Pindiville, J. Rodriguez, D. Schneberk, B. Wong, P. Vitello, Dead Zones in LX-17 and PBX 9502, *Propellants Explos. Pyrotech.* **2006**, *31*, 89, doi: 10.1002/prep.200600014.

[21] P. C. Souers, A. Hernandez, C. Cabacungan, R. Garza, L. Lauderbach, S.-B. Liao, P. Vitello, Air Gaps, Size Effect, and Corner-Turning in Ambient LX-17, *Propellants Explos. Pyrotech.*, **2009**, *34*, 32. doi: 10.1002/prep.200700232.

[22] R. W. Ashcraft, G. T. West, *Establishment of a Corner Turning Test Capability*, Report HMSMP-78-64, Mason and Hanger-Silas Mason Co., Inc., Amarillo, TX, USA **1978**. doi: 10.2172/6482021.

[23] L. G. Hill, T. R. Salyer, The Los Alamos Enhanced Corner-Turning (ECOT) Test, *16th International Symposium on Detonation*, Cambridge, MD, USA, July 15–20, **2018**.

[24] L. G. Hill, W. I. Seitz, C. A. Forest, H. H. Harry, *High Explosives Corner Turning Performance and the LANL Mushroom Test*, Report LA-UR-97-2509, Los Alamos National Laboratory, Los Alamos, NM, USA **1997**. doi: 10.1063/1.55675.

[25] E. V. Bukovsky, R. D. Chambers, E. L. Lee, K. T. Lorenz, The Shell Acceleration Experiment (SAX) - A Modern Corner-Turning Experiment, *16th International Detonation Symposium*, Cambridge, MD, USA, **2018**.

[26] C. M. Tarver, Modelling Shock Initiation and Detonation Divergence Tests on TATB-Based Explosives, *Propellants Explos. Pyrotech.* **1990**, *15*, 132. doi: 10.1002/prep.19900150404.

[27] B. E. Clements, X. Ma, W. L. R. P. J. Perry, C. L. Amstrong, P. Dickson, Shock Initiation Response of PBX 9502 Considering Rarefaction Wave Effects, *16th International Detonation Symposium*, July 15–20, **2018**, Cambridge, MD, USA.

[28] N. J. Whitworth, CREST modelling of PBX 9502 corner turning experiments at different initial temperatures, *J. Phys. Conf. Ser.* **2014**, *500*, 052050. doi: 10.1088/1742-6596/500/5/052050.

[29] Y.-y. He, Y. Long, Three Dimensional Numerical Simulation of Detonation Wave Propagation on Corner-Turning of Composition B, *J. Explos. Propellants*, **2007**, *30*, 63.

[30] Y. Han, Z.-h. Jiang, Y.-m. Huang, X.-P. Long, Numerical Simulation of Corner Turning of PBX-9404 and PBX-9502 Explosive, *J. Explos. Propellants*, **2011**, *34*, 30.

[31] L. E. Fried, P. C. Souers, *CHEETAH: A Next Generation Thermochemical Code*, Report UCRL-ID-117240, Lawrence Livermore National Laboratory, Livermore, CA, USA **1994**. doi: 10.2172/95184.

[32] P. C. Souers, P. Vitello, *Explosive Model Tarantula V1/JWL++ Calibration of LX-17: #2*, Report LLNL-TR-413518, Lawrence Livermore National Laboratory, Livermore, CA, USA **2009**. doi: 10.2172/956832.

[33] P. C. Souers, D. Haylett, P. Vitello, *TARANTULA 2011 in JWL++*, Report LLNL-TR-509132, Lawrence Livermore National Laboratory, Livermore, CA, USA **2011**. doi: 10.2172/1035287.

[34] P. C. Souers, J. W. Forbes, L. E. Fried, W. M. Howard, S. Anderson, S. Dawson, P. Vitello, R. Garza, Detonation Energies from the Cylinder Test and CHEETAH V3.0, *Propellants Explos. Pyrotech.* **2001**, *26*, 180.

[35] K. T. Lorenz, E. L. Lee, R. Chambers, A Simple and Rapid Evaluation of Explosive Performance - the Disc Acceleration Experiment, *Propellants Explos. Pyrotech.* **2015**, *40*, 95. doi: 10.1002/prep.201400081.

[36] B. D. Lambourn, H. R. James, The relation between reaction rate and shock strength - A possible second order improvement to the CREST reactive burn model, *17th Conference of the American Physical Society Topical Group on Shock Compression of Condensed Matter*, Chicago, Illinois (USA), June 26 – July 1, **2011**, AIP Conference proceedings 1426, p. 591. doi: 10.1063/1.3686348.

[37] P. C. Souers, P. Vitello, Initiation Pressure Thresholds from Three Sources, *Propellants Explos. Pyrotech.*, **2007**, *32*, 288. doi: 10.1002/prep.200700030.

[38] M. Gresshoff, Insensitive High Explosives Shock-to-Detonation Transition Criteria, *16th International Detonation Symposium*, July 15–20, **2018**, Cambridge, MD, USA.

[39] G. A. Levesque, P. Vitello, The Effect of Pore Morphology on Hot Spot Temperature, *Propellants Explos. Pyrotech.* **2015**, *40*, 303. doi: 10.1002/prep.201400184.

[40] F. D. Murnaghan, *Finite Deformation of an Elastic Solid*, New York: John Wiley and Sons, Inc., **1951**.

[41] F. Birch, The Effect of Pressure Upon the Elastic Parameters of Isotropic Solids, According to Murnaghan's Theory of Finite Strain, *J. Appl. Phys.*, **1938**, *9*, 279. doi: 10.1063/1.1710417.

[42] F. D. Murnaghan, Finite Deformations of an Elastic Solid, *Am. J. Math.*, **1937**, *59*, 235. doi: 10.2307/2371405.

[43] J. W. Kury, H. C. Hornig, E. L. Lee, J. L. McDonnel, D. L. Ornellas, M. Finger, F. M. Strange, M. L. Wilkins, Metal Acceleration by Chemical Explosives, *4th Symposium (International) on Detonation*, White Oak, Maryland, 12–15 October **1965**.

[44] E. L. Lee, H. C. Hornig, J. W. Kury, *Adiabatic Expansion of High Explosive Detonation Products*, Report UCRL-50422, Lawrence Radiation Laboratory, University of California, Livermore, CA USA **1968**. doi: 10.2172/4783904.

[45] E. L. Lee, H. C. Hornig, Equation of State of Detonation Product Gases, *12th Symposium (International) on Combustion*, **1969**, p. 493. doi: 10.1016/S0082-0784(69)80431–5.

[46] P. C. Souers, B. Wu, J. L. C. Haselman, *Detonation Equation of State at LLNL, 1995, Revision 3*, Report UCRL-ID-119262-Rev.3, Lawrence Livermore National Laboratory, Livermore, CA USA **1996**. doi: 10.2172/204120.

[47] G. Baudin and R. Serradeill, Review of Jones-Wilkins-Lee equation of state, *EPJ Web Conf.*, **2010**, *10*, 00021. doi: 10.1051/epjconf/20101000021.

[48] R. Menikoff, *JWL Equation of State*, Report LA-UR-15-29536, Los Alamos National Laboratory, Los Alamos, NM, USA **2017**. doi: 10.2172/1229709.

[49] D. J. Steinberg, S. G. Cochran, M. W. Guinan, A constitutive model for metals applicable at high-strain rate, *J. Appl. Phys.*, **1980**, *51*, 1498. doi: 10.1063/1.327799.

[50] A. L. Nichols, III, C. M. Tarver, A Statistical Hot Spot Reactive Flow Model for Shock Initiation and Detonation of Solid High Explosives, *12th International Detonation Symposium,* August 11–16, **2002**, San Diego, CA, USA.

[51] L. E. Fried, High Explosive Shock Initiation Model Based on Hot Spot Temperature, *16th International Detonation Symposium*, July 15-20, **2018**, Cambridge, MD, USA.

[52] T. M. Willey, D. M. Hoffmann, T. van Buuren, L. Lauderbah, R. H. M. A. Gee, G. E. Overturf, L. E. Fried, The Microstructure of TATB-Based Explosive Formulations During Temperature Cycling Using Ultra-Small-Angle X-Ray Scattering, *Propellants Explos. Pyrotech.* **2009**, *34*, 406. doi: 10.1002/prep.200800031.

[53] A. L. Nichols, III, J. Gambino, Toward a Morphologically Aware Detonation Model, *16th International Detonation Symposium*, July 15–20, **2018**, Cambridge, MD, USA.

---

NB: This is a color version of the official, published Open Access article available here:
https://onlinelibrary.wiley.com/doi/epdf/10.1002/prep.201900383